\documentclass[]{acmart}

\usepackage{xspace}

\newcommand{\Sys}{AICS\xspace}
\newcommand{\parlabel}[1]{\noindent{\bf #1.}}
\newcommand{\itparlabel}[1]{\vspace{0.5em}\noindent\textit{#1}.}
\newcommand{\proved}{\hfill $\square$ }

\copyrightyear{2023} 
\acmYear{2023} 
\setcopyright{acmlicensed}\acmConference[IoTDI '23]{International Conference on Internet-of-Things Design and Implementation}{May 9--12, 2023}{San Antonio, TX, USA}
\acmBooktitle{International Conference on Internet-of-Things Design and Implementation (IoTDI '23), May 9--12, 2023, San Antonio, TX, USA}
\acmPrice{15.00}
\acmDOI{10.1145/3576842.3582388}
\acmISBN{979-8-4007-0037-8/23/05}

\begin{document}

\title{Amalgamated Intermittent Computing Systems}

\author{Bashima Islam}
\affiliation{%
  \institution{Worcester Polytechnic Institute}
  \country{USA}
}
\email{bislam@wpi.edu}

\author{Yubo Luo}
\affiliation{%
  \institution{University of North Carolina Chapel Hill}
  \country{USA}
}
\email{yubo@cs.unc.edu}

\author{Shahriar Nirjon}
\affiliation{%
  \institution{University of North Carolina Chapel Hill}
  \country{USA}
}
\email{nirjon@cs.unc.edu}

\begin{teaserfigure}
    \includegraphics[width=\textwidth]{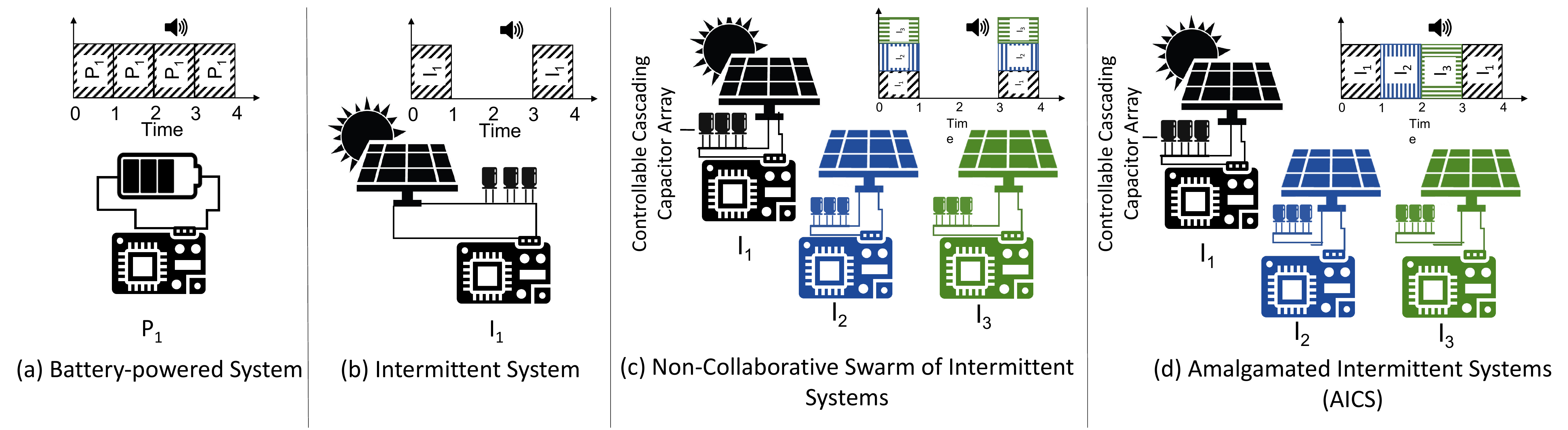}
    \caption{(a) A battery-powered system is always active and senses and processes all incoming events, (b) an intermittent system that misses the event at time 2 due to scarcity of energy, (c) Three individually performing intermittent systems without collaboration wake up at the same time and miss the event, and (d) the same three intermittent systems with appropriate duty-cycle amalgamate to sense and process all events within the deadline.}
    \label{fig:teaser}
    \Description{Overview Figure}
\end{teaserfigure}

\begin{abstract}
Intermittent computing systems undergo frequent power failure, hindering necessary data sample capture or timely on-device computation. These missing samples and deadlines limit the potential usage of intermittent computing systems in many time-sensitive and fault-tolerant applications. However, a group/swarm of intermittent nodes may amalgamate to sense and process all the samples by taking turns in waking up and extending their collective on-time. However, coordinating a swarm of intermittent computing nodes requires frequent and power-hungry communication, often infeasible with limited energy. Though previous works have shown promises when all intermittent nodes have access to the same amount of energy to harvest, work has yet to be looked into scenarios when the available energy distribution is different for each node.

The proposed \Sys framework provides an amalgamated intermittent computing system where each node schedules its wake-up schedules based on the duty cycle without communication overhead. We propose one offline tailored duty cycle selection method (Prime-Co-Prime), which schedules wake-up and sleep cycles for each node based on the measured energy to harvest for each node and the prior knowledge or estimation regarding the relative energy distribution. However, when the energy is variable, the problem is formulated as a Decentralized-Partially Observable Markov Decision Process (Dec-POMDP). Each node uses a group of heuristics to solve the Dec-POMDP and schedule its wake-up cycle.  
We conduct a real-world experiment by implementing a deep acoustic event classifier in three MSP430 microcontrollers. AICS successfully captures and processes 41.17\% more samples than a swarm of greedy intermittent computing systems while spending 69.7\% less time with multiple redundant active systems. Our simulation-based evaluations show a 35.73\%–54.40\% higher compute and process success rate with AICS than with state-of-the-art algorithms (including reinforcement learning).
\end{abstract}

\maketitle

\section{Introduction}

Harvesting ambient energy (solar, kinetic, thermal, and radio frequency) is considered a sustainable alternative to batteries~\cite{sudevalayam2010energy, merrett2016energy} for powering the next trillion Internet of Things (IoT)~\cite{sparks2017route}. However, ambient energy is often insufficient to power these devices directly due to lower energy availability than energy requirements. Hence, a battery-free system accumulates adequate energy in an energy buffer (e.g., super-capacitor) before operating. As a result, these devices go through frequent power-on and power-off periods and experience intermittency during execution. Thus, these systems are called \textit{intermittent systems}.

The intermittent system's advancement has occurred chiefly in programming language and system architecture, where the focus lies on the progress of instruction execution~\cite{balsamo2015hibernus, gobieski2019intelligence, yildirim2018ink}, memory consistency~\cite{hardin2018application}, and the development of emulators and debuggers~\cite{colin2016energy, colin2017energy}. Recent works have concentrated on intermittent systems' timing aspect to track time~\cite{dereliable, hester2016persistent, rahmati2012tardis}, finish data processing within a deadline~\cite{islam2020scheduling, islam2020zygarde}, discard stale data~\cite{hester2017timely, yildirim2018ink}, and increase energy buffer~\cite{colin2018reconfigurable}. Despite these commendable efforts, current intermittent nodes fail to sense or process events during a power failure, limiting their potential in continuous monitoring and fault-intolerant application domains. Another body of work~\cite{bhatti2014sensors} focuses on efficient energy harvesting and distribution techniques that provide continuous energy to wireless nodes using beamforming from a wall-powered source. However, such methods only apply when we have complete control over the energy source, which is often false. 

Three primary factors make event sensing and timely execution in a battery-free node challenging -- (1) \textit{dynamic available energy}, (2) \textit{asynchronous energy and event source}, and (3) \textit{sporadic occurrence of rapidly varying events}. 
Recent works have focused on formulating predictability~\cite{islam2020zygarde}, determining the optimal duty cycle for maximum wake-up time for a single node~\cite{fraternali2020aces}, and learning the distribution of an event's occurrence~\cite{smarton,luo2019spoton} to reduce the uncertainty of dynamic energy. However, these works have yet to have solutions against missing events during long power-off periods, where events refer to the occurrence of a specific phenomenon, e.g., bell ringing.

When the energy source and event are the same (e.g., solar-powered UV ray monitor), events only occur when power is available, guaranteeing the system's availability to capture even sporadic events. These are passive event sensing, where the presence of a source is the external trigger. However, the same can only be said for independent energy and target event source when they are aligned in time (e.g., solar-powered acoustic event detector)~\cite{bashima2019ondevice,lee2019intermittent}. These are active sensing systems that sense and compute through pooling. This asynchronous wake-up and event occurrence causes failure in sporadic events observation.

The practical significance of this loss of observability is dependent on what the sensors intend to sense. While a low sampling rate (1 sample/minute) is sufficient for indoor temperature monitoring, monitoring stochastic event-based data or rapidly varying environmental stimuli (e.g., sound event detection) requires more observation. As both dynamic energy and sporadic event are uncontrollable, despite all the efforts, battery-free nodes fail to sense and compute all samples to determine the events when the energy and event sources are uncorrelated.

Our key idea to decrease missing samples and unfinished computation is straightforward. Imagine a scenario where we aim to determine an alarm or bell ringing or what appliance is active in a room. The system uses a microphone to capture audio clips and computes on-device to determine the sound source with classification. Depending on the source type, it sends a notification.
A battery-powered system is always on and thus can capture data and identify when the bell rang at time unit 2 (Figure~\ref{fig:teaser}a). However, an intermittent system that activates once every three seconds misses capturing the event (Figure~\ref{fig:teaser}b). Even if a battery-free node captures the sample, it may fail to finish processing it in time, resulting in an undetected event. Multiple identical nodes are also redundant because they will all be active simultaneously (Figure~\ref{fig:teaser}c).

The opportunity lies in that only one of the many intermittent systems must capture and process the event successfully to identify if the bell rang. Thus, scheduling the wake-up and sleep cycle can amalgamate three identical intermittent systems with only one active node at any time. As a result, an amalgamated intermittent system captures and processes all the samples and can identify the bell ring at time unit 2 in Figure~\ref{fig:teaser}d. In summary, with sufficient intermittent systems working as an amalgamated system, we can emulate a battery-powered node when the power-off period does not exceed the energy buffer self-discharging time.

Though scheduling a swarm of nodes to work collaboratively is not a new problem~\cite{zheng2013survey, zhu2019broadcast}, these algorithms rely on active communication among the nodes. Frequent communication among multiple nodes is infeasible in intermittent computing systems, as communication is often the most power-hungry operation~\cite{gobieski2019intelligence}. Though zero-power passive communication~\cite{torrisi2020zero, geissdoerfer2021bootstrapping} is promising, these methods can communicate with only one node at a time, increasing communication time among all the nodes in a swarm. Moreover, while passive backscatter communication depends on an active receiver~\cite{torrisi2020zero} or transceivers~\cite{geissdoerfer2021bootstrapping}, they require precise (milliseconds or even nanoseconds) time synchronization to allow the sender and receiver to turn on the radio simultaneously. Bonito~\cite{geissdoerfer2022learning} proposes a learned wake-up schedule for effective communication, and Flync~\cite{geissdoerfer2021bootstrapping} provides an effective solution for neighbor discovery despite intermittency. However, these works only support one device and often require prior knowledge about the energy pattern. Moreover, required packet conflict recovery makes them insufficient for frequent communication across a swarm of systems. 

We propose \Sys, a framework that amalgamates a swarm of intermittent nodes for prolonging the collective power-on period without inter-node communication. \Sys allows each intermittent system to decide whether to wake up or go to sleep by predicting other nodes' behavior without communication. To increase successful capture and computing and maximize energy utilization, \Sys reduces the number of simultaneously active nodes when the power-off period is less than the self-discharging time.

Coalesced Intermittent Sensor (CIS)~\cite{majid2020continuous} is our closest related work which randomly wakes up each intermittent system to avoid concurrent active systems. However, it only considers the scenario where the same energy is available to every node of the swarm. Despite being an important scenario, it fails to provide a solution for a broad spectrum of stochastic scenarios that can happen in real-world. \Sys primarily focuses on the dynamic scenario where the available energy to each node is different. This scenario is more common as different swarm nodes will have access to variable amounts of energy from the same source due to their placement, distances, and other environmental factors, e.g., occlusion. 

Through \Sys, we make three technical contributions:

\begin{itemize} 
    \item We propose an algorithm (Prime-Co-Prime) that determines tailored sleep and wake-up duty cycles for each intermittent node using the relative energy distribution when the energy available to each system is static over time. However, each system has access to a different amount of energy. We also prove that the proposed algorithm requires a minimum number of swarm nodes.

    \item When the available energy varies over time and is also different at each node, we provide an online tailored duty cycle selection method by formulating a Decentralized Partially Observable Markov Decision Process (Dec-POMDP). Due to the high computation overhead to solve a Dec-POMDP, we extend the Prime-Co-Prime algorithm to define duty cycles as states and provide energy-efficient heuristics to allow offline transition among these states based on the measured energy at each node. These states provide the following action by choosing the correct and diligent duty cycles.  
\end{itemize}

To understand the feasibility of our approach, we develop solar and RF-powered custom prototypes with an MSP430FR5994~\cite{msp430} microcontroller and controllable cascading capacitor array. We deploy a deep learning-based acoustic event detector that can determine the occurrence of events and classify them. We evaluate with real-world and simulation-based experiments and compare \Sys with an oracle, a duty cycle approach, a reinforcement learning-based approach (ACES~\cite{fraternali2020aces}), and a greedy approach. \Sys, on average, achieves 54.40\%, and 35.73\% higher capture and process success rate than a swarm of nodes with greedy, and ACES, respectively. In addition, \Sys achieves a 41.17\% higher capture and process success rate and spends 69.7\% less time with multiple active nodes than a swarm of greedy battery-free nodes in real-world experiments.
\section{Amalgamated Intermittent Computing Systems}
An amalgamated intermittent computing system (\Sys) is a swarm of $N$ intermittent nodes where $N>1$. Each node has the same group of sensors to capture data and performs the same computational steps to process them. Intermittent computing nodes go through wake-up and sleep cycles due to the unavailability of continuous and sufficient energy. However, the availability of energy guides these on-off cycles. In other words, whenever a node harvests sufficient energy to operate, it wakes up. A few previous works~\cite{islam2020scheduling} have shown the benefit of waking up only when there is sufficient energy. Instead of waking a node up whenever there is energy, \Sys provides duty cycles to each node to keep the other nodes asleep when one node is active. This way, one of the other nodes can be active after the active node consumes all its energy and increases the active time. For example, suppose a node has 1 second operating time with the available energy and requires two additional seconds to accumulate sufficient energy to wake up. In that case, the node has a duty cycle of 3 seconds. If two other nodes can be active during those two seconds, the active time of the system gets a three times increase. A fixed duty cycle is sufficient if all nodes have access to the same available energy (like CIS). However, in most real-world scenarios, available energy at nodes of a swarm varies, and a fixed duty cycle is insufficient. \Sys utilizes two tailored duty cycle selection methods described in Section~\ref{sec:algo}, and allows the nodes to amalgamate and increase their collective active time and processing capability. 

These duty cycles can be pre-defined during the compile time or can be selected from a pre-defined list of duty cycles during runtime. We found that whether compile time selection is sufficient or a runtime selection is needed depends on the available energy's characteristics. Thus, before exploring the proposed duty cycle selection method, we describe the energy characteristics and the assumptions we make in this paper. 


\begin{figure}
    \centering
    \includegraphics[width=0.45\textwidth]{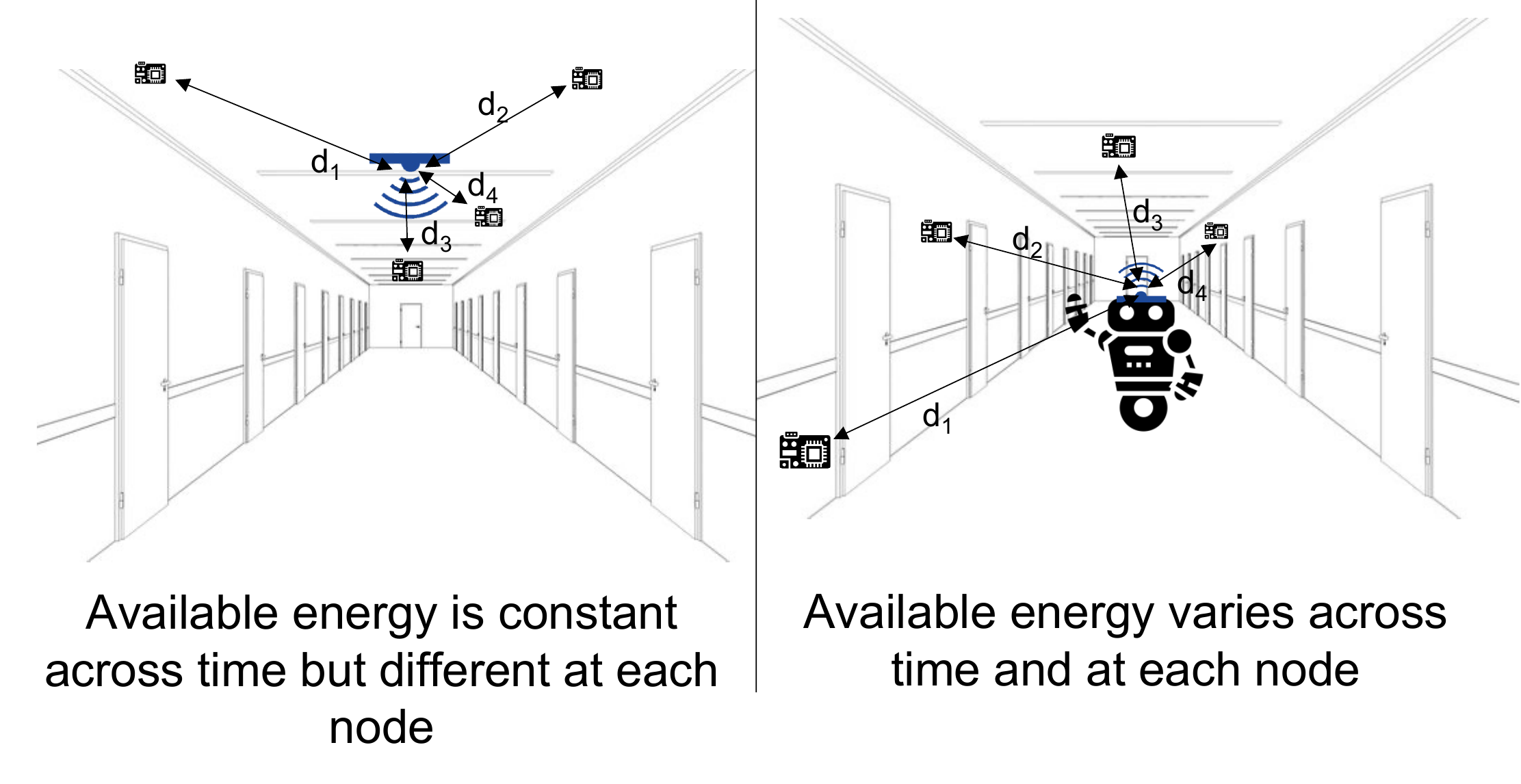}
    \caption{Different scenarios based on the energy availability}
    \label{fig:scenarios}
\end{figure}

\subsection{Energy Characteristics}
\parlabel{Available Energy}
To simplify the analysis, we quantize the available energy harvest into discrete energy levels by dividing the total energy at each time slot by unit energy. The constant time slot is empirically determined using a task's shortest execution time and the lowest energy consumption.

\parlabel{Energy Availability}
Depending on the stochasticity of the energy source, its availability varies over time. When the available energy is constant in time, a node gets unchanging energy, making it easier for the node to make future decisions. In Figure~\ref{fig:scenarios} (left), we show that when a node is harvesting from the RF transmitter mounted on the roof and there is no changing interference, the available energy to a node is unchanged. On the contrary, Figure~\ref{fig:scenarios} (right) shows a mobile RF transmitter mounted on a robot where the available energy at each node varies depending on the changing distance between the node and transmitter. 

The ubiquity of the available energy to all the nodes in a swarm can also be constant or variable. Figure 1 shows the two scenarios where each node has access to a diverse energy. This paper focuses explicitly on these scenarios. A swarm of RF-powered nodes at equidistant and line-in-sight for an RF transmitter have access to the same energy in time. This is a more straightforward case than the previous one because, unlike the previous case, each node can easily calculate the energy status of the other nodes. Previous works~\cite{majid2020continuous} have focused on the second scenario, and thus we focus on the more explored scenario where the available energy at each node varies.

\subsection{Assumptions}
We make the following assumptions in the paper. 
\begin{itemize}
    \item All the nodes are identical and have equal harvesting efficiency and energy consumption rate. Every node has the same components and executes the same procedure on the captured sample. This paper considers the worst-case scenario (lowest harvesting efficiency and highest consumption rate across all nodes) to ensure reliability. This assumption may result in suboptimal performance. However, our experiments found that the variation among five nodes with 100 iterations at each node is negligible. 

    \item The relative energy distribution among all the nodes is static and known as a priori. In other words, the nodes are not moving. 

    \item The scope of tasks is limited to those where all the nodes can capture any event. To illustrate, in the audio event detection in a room example in Section 1, all the nodes in various room positions can hear the same sound. On the contrary, tasks where the same data is not accessible to all nodes, e.g., vibration measurement at different walls, will be out of the scope of this paper. 

    \item We consider that the clocks of all nodes have the same start time. Over time these clocks can drift. Existing battery-free clocks have only $\pm$1 ms to -1 s of an error on 2.2 -- 2200 nF capacitors. The implementation of the proposed algorithm considers this drift and randomly adds counter-drifts to handle this. 
\end{itemize}

\section{Duty Cycle Selection}
\label{sec:algo}
Existing duty cycle algorithms use a fixed duty cycle for all the nodes in a system. When the available energy is constant over time and is equal to all the system nodes, a fixed duty cycle is feasible for amalgamated intermittent computing systems. Even without communication, the nodes can keep track of the energy budget of the other nodes as they have the same available energy. However, a node needs to know when the other nodes are active. Thus, by scheduling with a pre-defined duty cycle and staggered start time, an amalgamated system can ensure at least one active node at any given point in time without the additional computational overhead.

However, this work focuses on a more diverse scenario where the available energy varies at each node. We observe that \textit{a standard pre-defined duty cycle is not optimal even if the available energy is invariant over time but deviates at each node}. We formally proof it below. 

\itparlabel{Proof}
We prove this statement using contradiction. Let us assume that a standard pre-defined duty cycle is optimal when the energy source is constant over time but variable at each node. In such a scenario, the duty cycle is calculated with the lowest available energy to a node to ensure that at least one node is active. Taking the highest available energy to a node is not suitable due to the lack of guarantee and possible lack of sufficient energy to obey the defined duty cycle. However, the nodes with higher available energy will waste energy as their energy storage will get charged faster. If these nodes have small duty cycles, the total number of nodes required to ensure one active node at any time can be reduced. It contradicts our earlier assumption that a common pre-defined duty cycle is optimal. Thus, the common and pre-defined duty cycle is not optimal when available energy is consistent over time but differs across nodes. 
\proved

\subsection{Tailored Duty Cycle}
Thus, we explore the opportunity to define a tailored duty cycle for each node in the system by proposing a \textit{Prime-Co-Prime (PCP)} duty cycle selection algorithm. PCP provides the duty cycle based on the prime and co-prime numbers. Prime numbers refer to numbers greater than 1 and only have two factors, one and the number itself. Therefore, we will require at least one node for each prime number to have an active node at all times. If the lowest allowed duty cycle is 2 having additional nodes will increase the redundant active time where more than one is active at any time. However, in the real world, the lowest allowed duty cycle varies based on the available energy and capacitor size. To accommodate this, we exploit co-prime numbers of the allowed prime numbers. Two positive integers are called co-prime if and only if they have 1 as their only common factor. This algorithm is optimal and can ensure that at least one node is active at any time using the minimum number of nodes. 

\begin{figure}[!htb]
    \centering
    \includegraphics[width=0.45\textwidth]{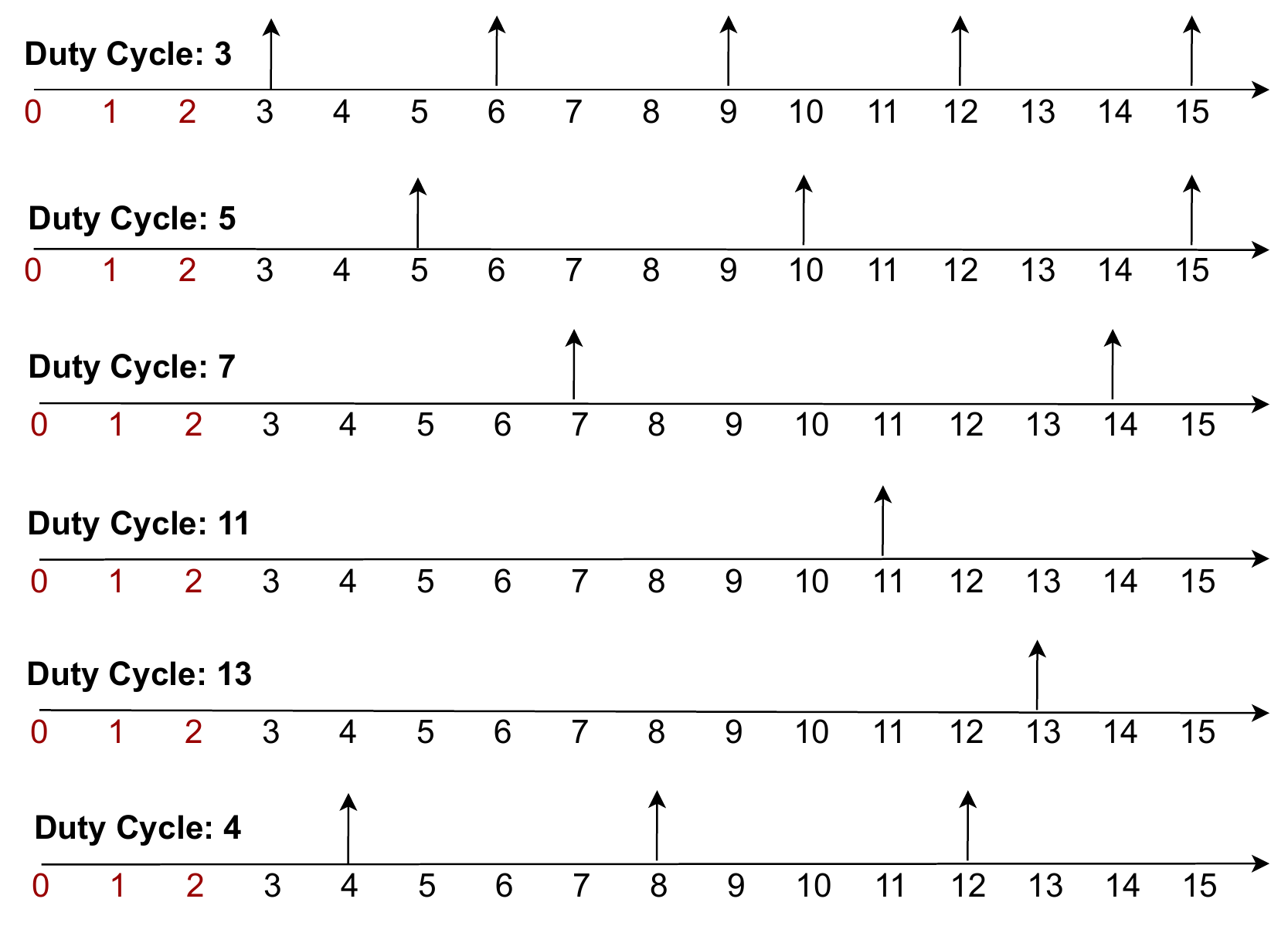}
    \caption{Prime-Co-Prime (PCP) with the lowest allowed duty cycle of 3 and hyper-period 15.}
    \label{fig:primecoprime}
\end{figure}

The Prime-Co-Prime duty cycle is determined offline during the compile time and is calculated in three steps. 

\begin{enumerate}
    \item First, we calculate the hyperperiod, T. The hyperperiod is the smallest interval after which the periodic patterns of all the tasks are repeated. In Figure~\ref{fig:primecoprime}, the hyperperiod is 15. 
    \item Next, we take all the prime numbers, P, where P<Q and T is the lowest duty cycle possible at any location. The example in Figure~\ref{fig:primecoprime} has the lowest allowed duty cycle of 3, and the prime numbers are 3, 5, 7, 11, and 13. 
    \item Finally, we use the Sieve of Eratosthenes to determine the rest of the duty cycles larger than T and are not divisible by P. The Sieve of Eratosthenes is an iterative process that is often used to find the prime numbers until a limit. We use it for now only to identify the prime and co-prime numbers of the prime list. In Figure~\ref{fig:primecoprime}, we observe that the co-prime to the prime list of the previous step is 4, which ensures that at least one of the 6 nodes are active at any time. 
\end{enumerate}

We formally prove that -- 
\textit{when the energy is constant in time and variable at each node, the minimum number of intermittent nodes required to have at least one node active at any time within a hyperperiod (T) can be given by a total number of primes smaller than or equal to T.} 
Here, the smallest allowed duty cycle is 2.

\itparlabel{Proof}
We prove this theorem by contradiction. Let us assume that for every prime number smaller than or equal to hyperperiod if there exists an intermittent node with a prime duty cycle. Thus, there exist time instances where no node is active.
Each number smaller than or equal to T can be either prime or non-prime. Moreover, as the nodes are co-prime, each duty cycle will be unique. If the time instance represents a prime number, then there is an intermittent node with a prime duty cycle, and thus precisely one node will wake up at those points. 

For a non-prime number, we must have $q=mn$, where $1<m,n<q$. By induction, as $m$ and $n$ are smaller than $q$, they must each be a product of primes. Therefore, $q$ is also a product of prime. Therefore there will be $p$ actives nodes at $q$ time where $p$ is the number of unique primes factors of $q$. 

It contradicts our assumption.
\proved

\subsection{Online Tailored Duty Cycle}
When the available energy varies with time and node, having a pre-defined tailored duty cycle is not sufficient. As the available energy changes over time, the possible duty cycle for each node also changes, and the pre-defined duty cycles can not be achieved. To address this, we form this problem as a Decentralized Partially Observable Markov Decision Process (Dec-POMDP), where there is no communication between the agents (nodes), and the state (energy status) is partially known (each node knows its harvestable energy). A markovian decision process  (MDP) formalizes sequential decision-making where a state's actions depend on immediate rewards and subsequent states. In our scenario, each node's decision depends not only on whether it successfully captured and processed an event but also on the subsequent energy status to allow future success in the duty cycle. 

An agent cannot directly observe the underlying state in a Partially Observable Markov Decision Process (POMDP), which is suitable for our scenario as a node can not directly observe the energy status of the other nodes. Finally, in a Decentralized Partially Observable Markov Decision Process (Dec-POMDP), there is no communication among the nodes, and decentralized decisions are made by each node individually. In the amalgamated intermittent computing system, there is no communication among the nodes; thus, we formulate the problem as a Dec-POMDP. Each node will determine its duty cycle online at the available energy change to this node. 

A Dec-POMDP is represented by a tuple $\{N, S, A, T, R, \Omega, O, h, b_0\}$. Here, $N = \{ 1, .., n\}$ is the set of n agents, $S$ is the finite set of states $s$, $A$ is the set of joint actions $a = \{a_1, .., a_n\}$, $T$ is the transition function that specifies $Pr(s_{t+1}|s_t, a_t)$, $R(s, a)$ is the immediate reward function, $\Omega$ is the set of joint observations $o = \{o_1, ..., o_n\}$, $O$ is the observation function: $Pr(o_{t+1} | a_t, s_{t+1})$, $h$ is the horizon of the problem, and $b_0 \in \Delta(S)$ is the initial state distribution at time $t=0$.

Dec-POMDP aims to find an optimal joint policy $\pi*$ that maximizes rewards' expected sum (over time-step). We do not use the discounted summation of the rewards as performing a task sooner does not benefit our goal. Instead, having at least one intermittent node active all the time is more critical.

The main difference between multiagent MDP and Dec-POMDP frameworks is that the joint policy is decentralized. $\pi*$ is a tuple $\{\pi_i, ..., pi_n\}$ where the individual policy $\pi_i$ of every agent $i$ maps individual observations histories $o_{i, t} = \{o_{i, 1}, ..., o_{i, t}$ to action $\pi_i(o_{i,t} = a_{i,t})$.

\parlabel{Reward Function, R} The immediate reward function is crucial for solving a Dec-POMDP. A reward function can be positive, negative, or even a combination. The positive reward is more straightforward to calculate as any time a node successfully captures and processes an event, we assign a positive reward. However, a negative reward is more challenging to determine as the ground truth of whether an event went missing is not available to a node. We exploit application-specific characteristics to define the negative reward. To illustrate, in the audio event detection example mentioned above, if a node captures an event from the first time after waking up, it can assume that the audio event started when the node was asleep before waking up and missed the event. Thus, we will assign a negative reward. 

When the application involves monitoring and maintaining a continuous variable, we can exploit that variable to assign a negative reward. For example, if a bell ringing indicates the end of a manufacturing cycle, the manufacturing machines will reach a steady state for cool-down. An intermittent node can monitor the movement and temperature of the machines during each wake-up to identify whether it has missed an event and assign a negative reward.

Despite being a suitable solution, DEC-POMDP is unsuitable for an intermittent node due to its \textit{NEXP completeness}. 
As the energy at each node is non-uniform, the number of states can be hard to bound. To reduce the complexity, we use the duty cycles derived from the Prime-Co-Prime algorithm to deduce states. To illustrate, in the example of Figure~\ref{fig:primecoprime} the states-set includes all permutations of $\{3, 4, 5, 7, 11, 13\}$. Therefore, all nodes in the same state will have at least one active node. Next, we design the state diagram using knowledge from an approximate offline solution of the DEC-POMDP. However, along with additional computational overhead, using the DEC-POMDP transition function requires observation history, which introduces additional memory overhead and is unsuitable for intermittent nodes. 

Thus we provide two lightweight heuristics to transition between states. As a node does not know the state of other nodes, these heuristics are suboptimal. The following are some proposed heuristics:

\vspace{5pt}
\parlabel{Randomized Binary Search (RBS)}
Upon experiencing changing available energy, a node randomly chooses a state using a variant of the binary search method. In this method, instead of selecting the middle point, a node selects a random position between the range.
We use randomization instead of the traditional binary search method to ensure that intermittent nodes can recover from scenarios where all of them have selected the wrong state sets. On the contrary, this solution may also force nodes to choose the wrong state sets when they were already at the correct one. 

\vspace{5pt}
\parlabel{Suboptimal Reinforcement Learning (SRL)}
In this heuristic, each node maintains a local transition matrix where sleep or wake are the actions and uses it to make individual decisions. A node exploits feedback to update the offline transition table whenever feedback is available. Here, each node maintains a local Q-Table where choosing the new duty cycle is the action. Each node makes individual decisions from the local Q-table and updates it using the feedback from the continuous variable. This heuristic is suitable for systems that monitor and maintain a continuous variable.


\section{Performance Characterization}This section describes the baseline algorithm and performance metric for the evaluation. 

\subsection{Baseline Algorithms}
We compare \Sys against four baseline algorithms -- oracle, greedy, pre-defined duty cycle, and ACES~\cite{fraternali2020aces}. 
For the greedy and ACES algorithms, we consider two variants -- (1) a single node and (2) a swarm of nodes, where wake-up and sleep scheduling happens locally in each node. We leave out simpler algorithms, e.g., round-robin, because a node does not know whether another node has finished executing. To preserve the evaluation's integrity, every algorithm with a swarm of nodes has the same number of nodes. 

\begin{itemize}
    \item \textit{Oracle Algorithm (\textbf{ORCL}).}
    The oracle solution has perfect and prior knowledge of all nodes' energy status without additional computation or memory overhead. At any point in time, the node with the maximum total available and stored energy (in the energy buffer) remains awake if the maximum total energy is more than or equal to the minimum operational energy. 

    \item\textit{Greedy Algorithm (\textbf{GRDY}).}
    In the greedy scheduler, a node wakes up whenever it accumulates enough energy to turn on the microcontroller and executes as long as energy storage has sufficient energy before going to sleep or low-power mode.

    \item \textit{Duty Cycle Algorithm (\textbf{DC}).}
    This algorithm has fixed pre-defined duty cycles for all nodes. However, it starts the nodes at different start times to keep at least one active node at any given time. If $t_h$ is the harvesting time and $t_e$ is the execution time, the duty cycle of the nodes is $t_e + t_h$ (if $t_h$ is divisible by $t_e$) or $t_e + t_h + 1$ (if $t_h$ is not divisible by $t_e$). An offset, $(n-1)t_e$,  determines the start time of the nodes; here, $n$ is the number of the node. 

    \item \textit{Automatic Configuration of Energy Harvesting Sensors (\textbf{ACES}).}
    ACES uses reinforcement learning to maximize each battery-free node sensing performance. Every 15 minutes, each node uses Q-learning to choose between four duty-cycle periods -- 15 seconds, 1 minute, 5 minutes, and 15 minutes.

\end{itemize}

\subsection{Performance Metric}
\label{sec:metric}
We evaluate the performance of our proposed method using two evaluation metrics. 

\begin{itemize}
    \item \textit{Capture and Process Success Rate \textbf{($\zeta$)}.}
    Real-time systems define \textit{schedulability} as a scheduler's ability to schedule all processing tasks in a task set. An amalgamated intermittent system's performance depends on more than just finishing processing within the deadline. It is crucial to consider the success of capturing events. CIS proposes \textit{availability} which refers to when at least one node is active and can capture an event. Nevertheless, it fails to consider finishing a process within the deadline. 

    Hence, we propose the capture and process success rate, defined as $\zeta = \dfrac{|N_C \cap N_P|}{|N|}$, where $N_C$ is the set of successfully captured events, $N_P$ is the set of successfully processed events within the deadline, and $N$ is the set of all events. 
    Given a scheduling algorithm that schedules the wake-up and sleep of intermittent nodes, if all energy and event combinations can be captured and processed within the deadline, we achieve 100\% $\zeta$. If a scheduler can capture and finish processing 50\% of the combined set events, the scheduler has 50\% $\zeta$. If a node only captures the event and fails to process the sample within the deadline, it does not contribute to $\zeta$.
    
    \item \textit{Redundant Active Time \textbf{($\Gamma$)}.}
    As multiple active nodes waste resources, we want only one node to be active at any time. Thus, we measure the redundant active time ($\Gamma$), the percentage of the total time more than one node is active. We define $\Gamma = \dfrac{T - T_{N_1}}{T}\times100$, where $T$ is the total time, and $T_{N_1}$ is when only one node is active. To illustrate, if a single node is active at any time, the redundant active time is 0\%. If only one node is active 30\% of the time, the redundant active time is 70\%. 
    
\end{itemize}
\section{Simulation Based Evaluation}
We first evaluate with simulation by observing 10,000 scenarios (parameter combinations) for an exhaustive evaluation. The following section describes our real-world evaluation. 

\subsection{Source, Event and Taskset}
This section describes the synthetic energy source, event, and data processing pipeline. 

\parlabel{Synthetic Energy Source Generation}
For the simulation, we generate synthetic energy traces guided by real-world measurements. The synthetic traces reflect two types of energy sources -- solar and radio-frequency (RF). For solar, we have measured the available energy in indoor and outdoor scenarios at different lighting conditions (e.g., sunny vs. rainy day). For RF, we measured energy loss for various distances and obstacles (e.g., wood, metal, wall, human) between the RF receiver and transmitter operating at 900Hz. We further utilize different RF energy path loss models and motion trajectories. We move the source in these trajectories at a wide range of speed to change its distance from the nodes and line of sight parameters.

These energy traces ensures the presence of three energy conditions, where the available energy is (1) greater than, (2) smaller than, or (3) equal to the required operating energy. When the available energy is more than the required energy, the system stores residual energy for the future when there is insufficient available energy. The energy storage capacity is randomly selected from valid capacitor sizes ranging from 2.2 nF to 1 F. This limited storage size ensures an energy overflow reflecting real-world energy constraints. In the beginning, $t = 0$, all energy storage is empty. 

We simulate two main energy availability scenarios when the available energy -- (1) is constant over time but is different at each node and (2) is variable across time and different for every node.

\parlabel{Synthetic Event Generation}
We aims to successfully monitor stochastic events or rapidly varying environmental stimuli (e.g., a bell ring, a car passing a pedestrian). Thus the synthetic event dataset contains 1,000 sporadic events where the range of period (the minimum difference between two consecutive events) and event duration are input to the random event generator. We avoid simultaneous event occurrence by making the period the upper bound for the event duration.

\parlabel{Synthetic Computing Task Generation}
All intermittent nodes perform the same task when they capture the event. We choose the task duration and energy consumption from a bin of 15 tasks -- RSA encryption, signal processing-based speaker detection, KNN-based audio classification, DNN-based audio classification, DNN-based keyword spotting, DNN-based image classification, Decision Tree based image classification, temperature anomaly detection with local outlier factor (LOF), signal processing-based shape detection, activity recognition, cuckoo filtering, blowfish encryption, bit count, DNN-based visual wake word, and DNN-based image recognition. We measure these tasks' runtime and energy consumption executing on an MSP430FR5994 microcontroller. Additional sensors consume more energy to operate; hence, we also consider the power consumption during sensor activation. During one iteration, this duration is constant and the same for all the nodes.

\parlabel{Simulating Clock Drift and Charging Time}
We consider the clock drift time and time to charge the capacitor for a realistic simulation. We emulate the clock drift of a battery-free timekeeper~\cite{dereliable} by considering its error bound. We design a software-controlled cascading capacitor array (Section~\ref{sec:hardware}) to reduce the capacitor charging time. We measure the time to charge the capacitor array at different energy rates to simulate the capacitor charging time and utilize these measurements.

\parlabel{Node Configuration}
We rely on the period of events to determine the total number of nodes during each simulation iteration. We first calculate the events' hyperperiod and the total number of primes and co-primes, $N$, within that hyperperiod. The simulator then simulates $N$ number of nodes. For fairness, all algorithms are evaluated on N nodes for each iteration. 

\subsection{Performance Analysis}
This section analyzes the performance of \Sys with various duty cycle selection algorithms.

\begin{figure}
    \centering
    \includegraphics[width=0.45\textwidth]{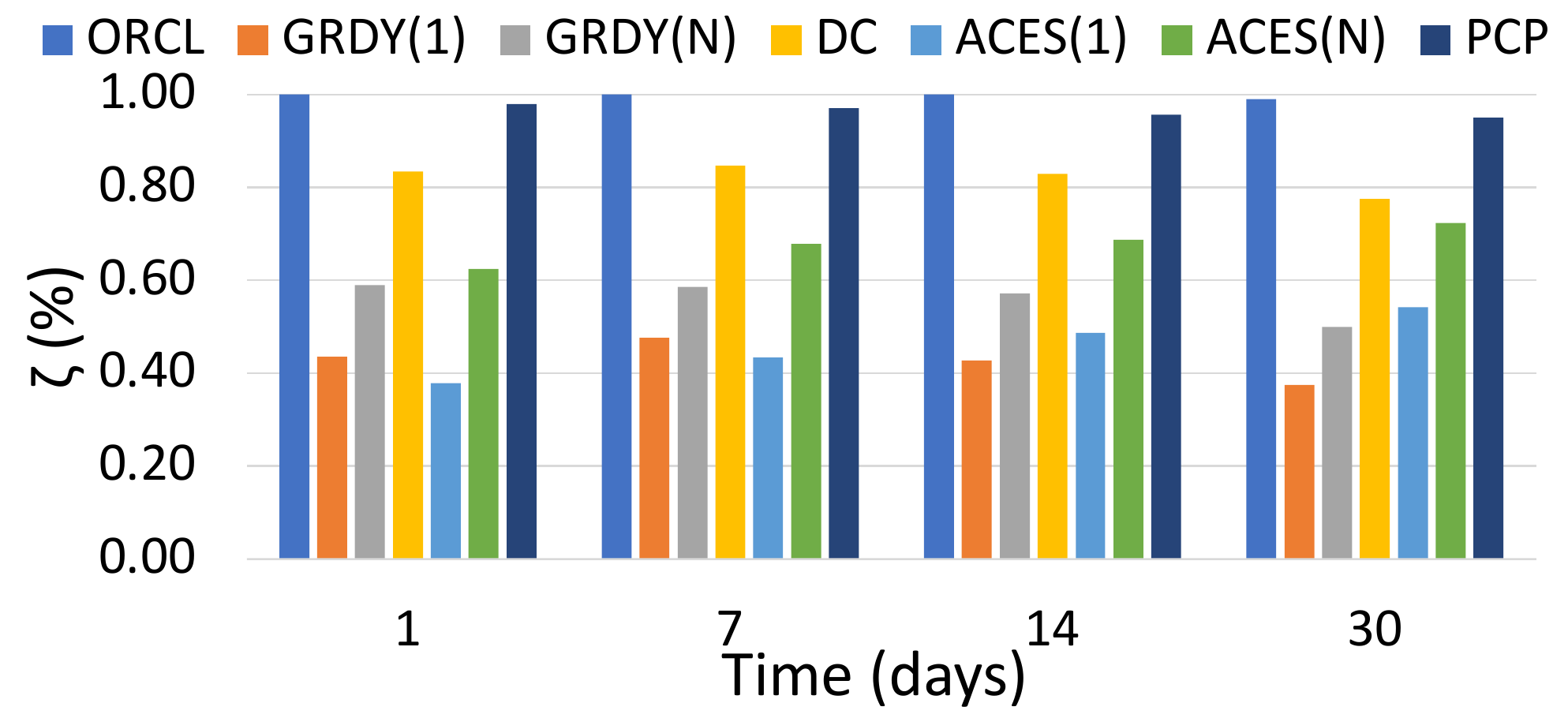}
    \caption{The average capture and process success rate for various algorithms when the available energy is constant over time but varies for each node. The fixed duty cycle algorithm can successfully capture and process 82\% of events. PCP determines custom and required duty cycles for each node and performs similarly to the oracle algorithm.}
    \label{fig:const_unbal_CPSR}
\end{figure}

\parlabel{Available Energy is Constant over Time but is Different for Each Node}
Figure~\ref{fig:const_unbal_CPSR} demonstrates different algorithms' capture and process success rate ($\zeta$) when the available energy is invariant over time but differs at each node. The X-axis of Figure~\ref{fig:const_unbal_CPSR}  represents the number of days the nodes were operating, and the Y-axis is the capture and process success rate, $\zeta$, defined in Section~\ref{sec:metric}. 

The Oracle (ORCL) algorithm can successfully capture and process 99.97\% of the data on average. Some $\zeta$ drop happens due to the failure to process all the sensed tasks rather than miss the events. The mean $\zeta$ of single-node operating greedily, GRDY(1), is 43\%, and it randomly varies with time as the event occurrence and energy availability are unsynchronized. To see whether having just N nodes increases the $\zeta$, we investigate N nodes executing the greedy algorithm, GRDY(N). We see a 13\% mean $\zeta$ increase in the system with a single node. Even though these nodes are not synchronized and lack a collective goal, they have a variable amount of energy available. Thus, they require a different amount of time to accumulate enough energy to turn on. It organically creates some desynchronization extending the $\zeta$.  

Next, we observe that when N nodes have the same pre-defined duty cycle, the $\zeta$ is 82\% on average. This performance is lower than the Oracle because each node requires various amounts of time to recharge the energy storage, and some nodes thus can not be active when they were supposed to be active. We also find that this performance drops over time as the coarse time synchronization among the nodes drifts. Note that all algorithms have the same number of nodes (besides the single cases) determined by the Prime-Co-Prime algorithm for a fair comparison. This is fair, as having more than the required nodes does not negatively impact the capture and process success rate. DC can achieve the $\zeta$ of the Oracle if it considers the minimum available harvestable energy to determine the duty cycle and the number of nodes at the cost of a higher number of nodes.

On the other hand, though a single node with the ACES algorithm, ACES(1), performs worse than the greedy algorithm on the first day, over time, it learns the energy distribution and increases the performance by 16\% in 30 days. A swarm of nodes executing ACES, ACES(N),  has higher $\zeta$ as each node randomly updates its action (duty-cycle) using reinforcement learning. It captures and processes 68\% of the events on average. 

The node-specific duty cycles determined by the Prime-Co-Prime (PCP) algorithm capture and process 96\% of the events on average. Though we observe a 3\% performance drop over the 30 days, in the first 14 days, the performance drop is only 2\%. The clock drift is responsible for this error, and we observed that without the random counter drifts, the performance dropped over 30 days become 8\%.


\begin{figure}[!htb]
    \centering
    \includegraphics[width=0.45\textwidth]{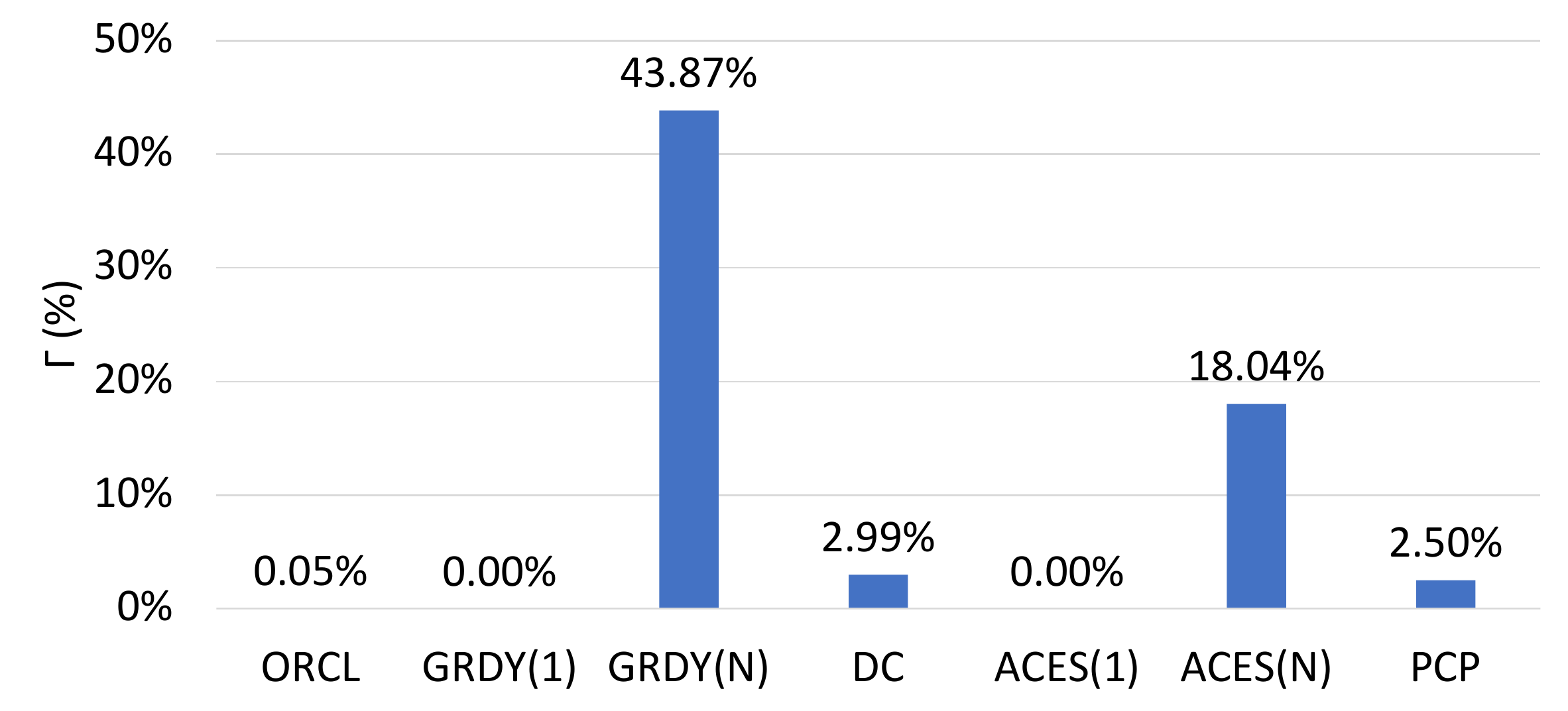}
    \caption{Average redundant active time for various algorithms when the available energy is constant over time but varies for each node. The duty cycle and Prime-Co-Prime algorithm have some redundant active time due to the clock drift.}
    \label{fig:const_unbal_red}
\end{figure}

In Figure~\ref{fig:const_unbal_red}, X-axis is various algorithms, and Y-axis is the redundant active time in percentage. The single-node algorithms --GRDY(1) and ACES(1)--have no redundant active time as they have only one node in the system. The Oracle has less than 1\% redundant active time, while the greedy swarm of nodes has 43.87\% redundant active time. There is no collaborative notion for the greedy swarm, and multiple nodes are often active simultaneously. The pre-defined fixed duty cycle algorithm has less than 3\% of redundant active time, often caused by clock drifts. Despite the lower redundant active time, this algorithm has a lower $\zeta$ because this system has 22.5\% idle time (no node was on) and failed to process 7.25\% of the captured events. ACES(N) also have higher redundant active time because the nodes have no collaborative notion. However, it is better than the greedy swarm as each node is using reinforcement learning to improve its performance, contributing to the comparatively lower redundant active time. Finally, the system with PCP has only 2.5\% redundant active time because a time value can be multiplication of more than one prime or co-prime numbers making all those nodes active at that time. PCP has only 1.89\% idle time and fails to process 7.28\% of captured samples, which contributes to reducing $\zeta$.


\begin{figure}[!htb]
    \centering
    \includegraphics[width=0.45\textwidth]{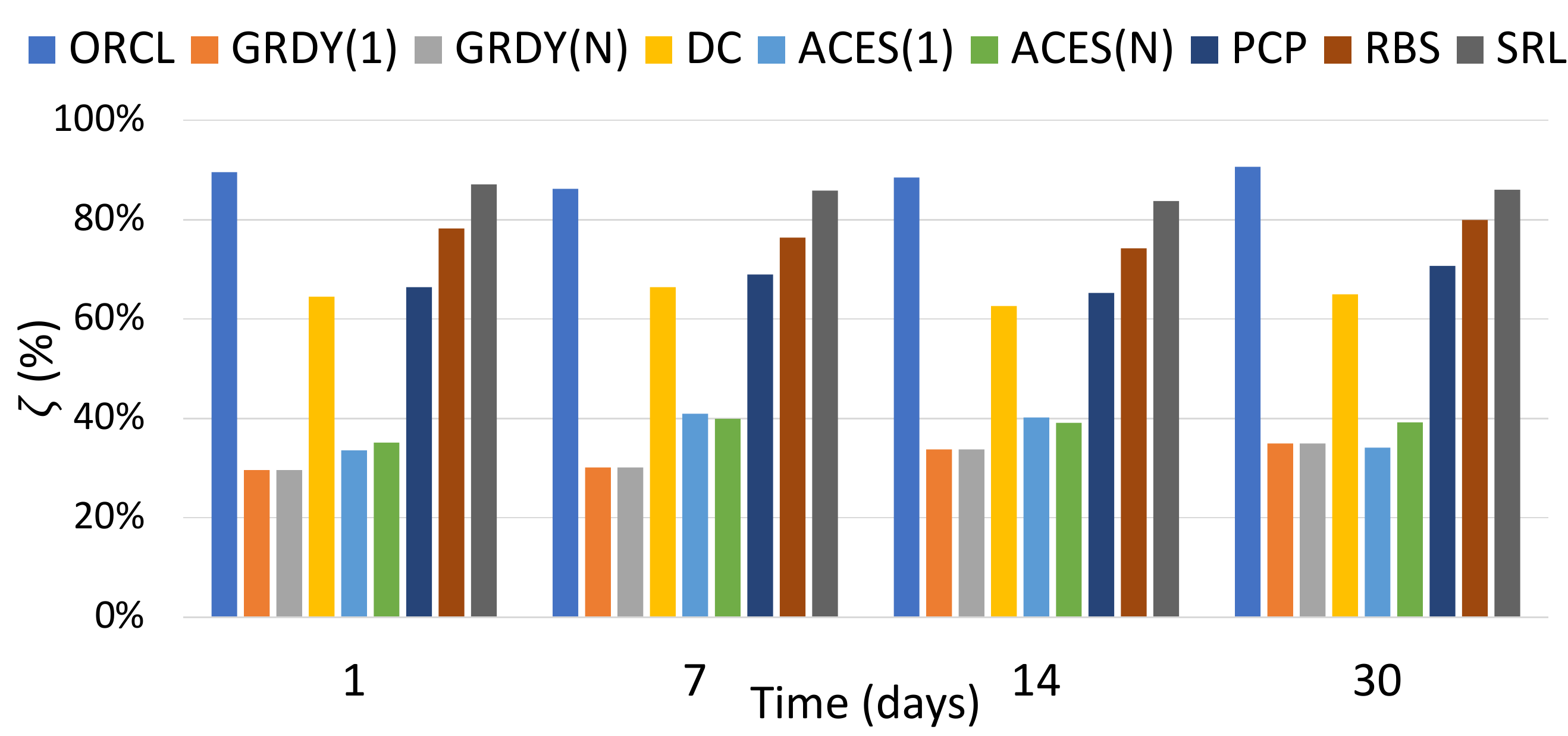}
    \caption{The capture and process success rate for various algorithms when the available energy varies over time and at each node.}
    \label{fig:var_unbal_CPSR}
\end{figure}

\parlabel{Available Energy Varies over Time and at Each Node}
Next, we look at the most exciting and practical scenario where the available energy varies over time, and every node gets a different amount of energy. Figure~\ref{fig:var_unbal_CPSR} shows different algorithms' capture and process success rate over 30 days. We observe that even the Oracle does not have a 100\% $\zeta$. The unknown variability of available energy causes energy scarcity for nodes, and thus, the system fails to process 9.34\% of captured events contributing to the total 11.27\% $\zeta$ drop. This variation is also responsible for the performance drop of the PCP algorithm. The PCP algorithm determines the tailored duty cycle for each node at the compile time. However, as the energy source varies over time, these duty cycles can not be maintained, resulting in long (30.72\%) idle time. 
With the Randomized Binary Search (RBS) heuristic of the DEC-POMDP formulation, we can successfully capture and process 10\% more events than the PCP. The Suboptimal Reinforcement Learning (SRL) based heuristic achieves higher $\zeta$ and can capture and process only 3\% fewer events than the Oracle. 
Using the prior knowledge about the relation of the available energy among the nodes and continuously learning from the feedback, SRL achieves this significant improvement. 

\begin{figure}[!htb]
    \centering
    \includegraphics[width=0.45\textwidth]{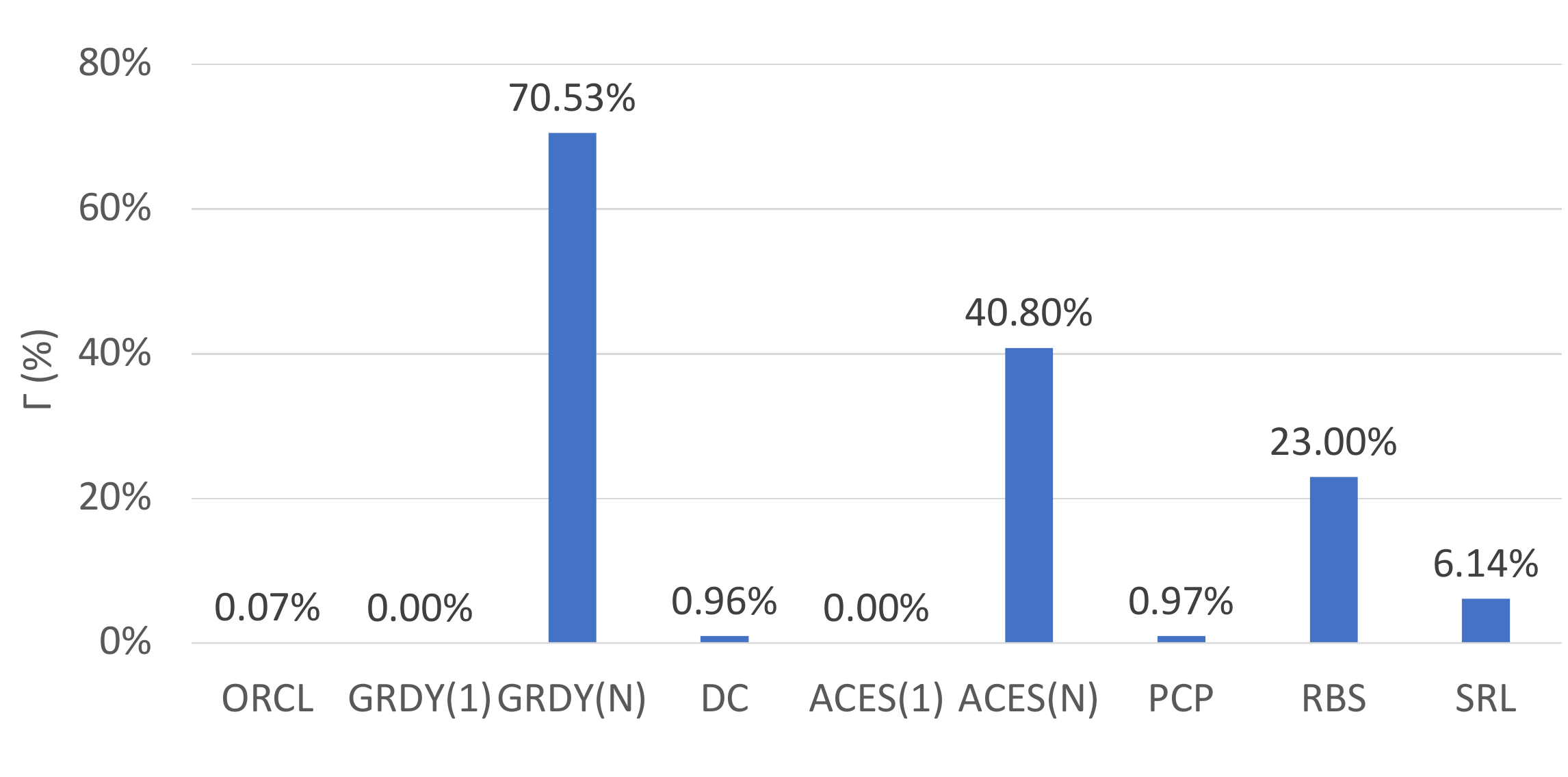}
    \caption{Redundant active time for various algorithms when the available energy varies over time and at each node. The swarm of greedy and ACES experiences increased redundant active time as the energy to harvest varies over time.}
    \label{fig:var_unbal_red}
\end{figure}

Figure~\ref{fig:var_unbal_red} shows the redundant active time when the energy varies over time and across nodes. Similar to the previous scenario, we observe that the Oracle and PCP have negligible redundant active time, while a swarm of greedy and ACES has higher redundant active time. Despite having a lower redundant active time, PCP's lower capture and process success rate is due to the 23.32\% idle time where the nodes fail to wake up despite their duty cycle. Though the RBS heuristic has 23\% redundant active time, it has a higher $\zeta$ than the PCP algorithm as it dynamically changes each node duty cycle. Using the prior knowledge about the relation of the available energy among the nodes and continuously learning from the feedback, SRL achieves even lower redundant active time. 
\section{Real world Evaluation}
In this section, we evaluate  AICS in the real world in two energy harvesting scenarios by developing a solar and RF-powered acoustic event detector using MSP430 microcontroller-based nodes. We use energy and event traces from real-world datasets for a fair comparison. We design prototypes of intermittent computing platforms with software-controlled cascading capacitor arrays described in Section~\ref{sec:hardware}.

\begin{figure}[!htb]
    \centering
    \includegraphics[width=0.45\textwidth]{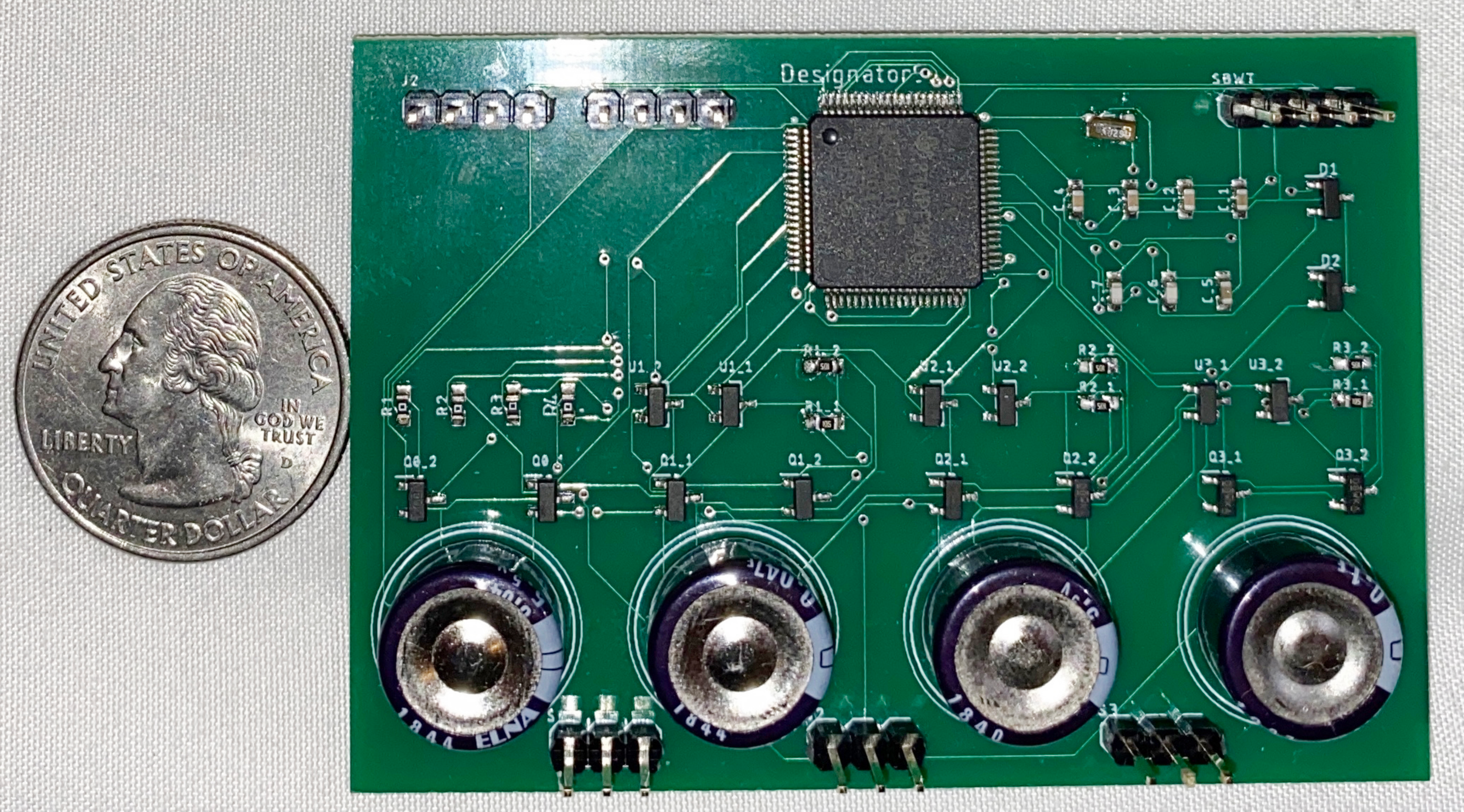}
    \caption{Custom PCB board with software-controlled cascading capacitor array}
    \label{fig:falinks_pcb}
\end{figure}


\subsection{Implementation and Experimental Setup}
\label{sec:hardware}

\parlabel{Hardware Platform}
We design an MSP430FR994-based prototype shown in Figure~\ref{fig:falinks_pcb} with a unique software-controlled cascading capacitor array that can dynamically customize the number and size of capacitors depending on the requirement. This capacitor array reduces the charging time and supports a broader range of applications.

\begin{figure}[!htb]
    \centering
    \includegraphics[width=0.45\textwidth]{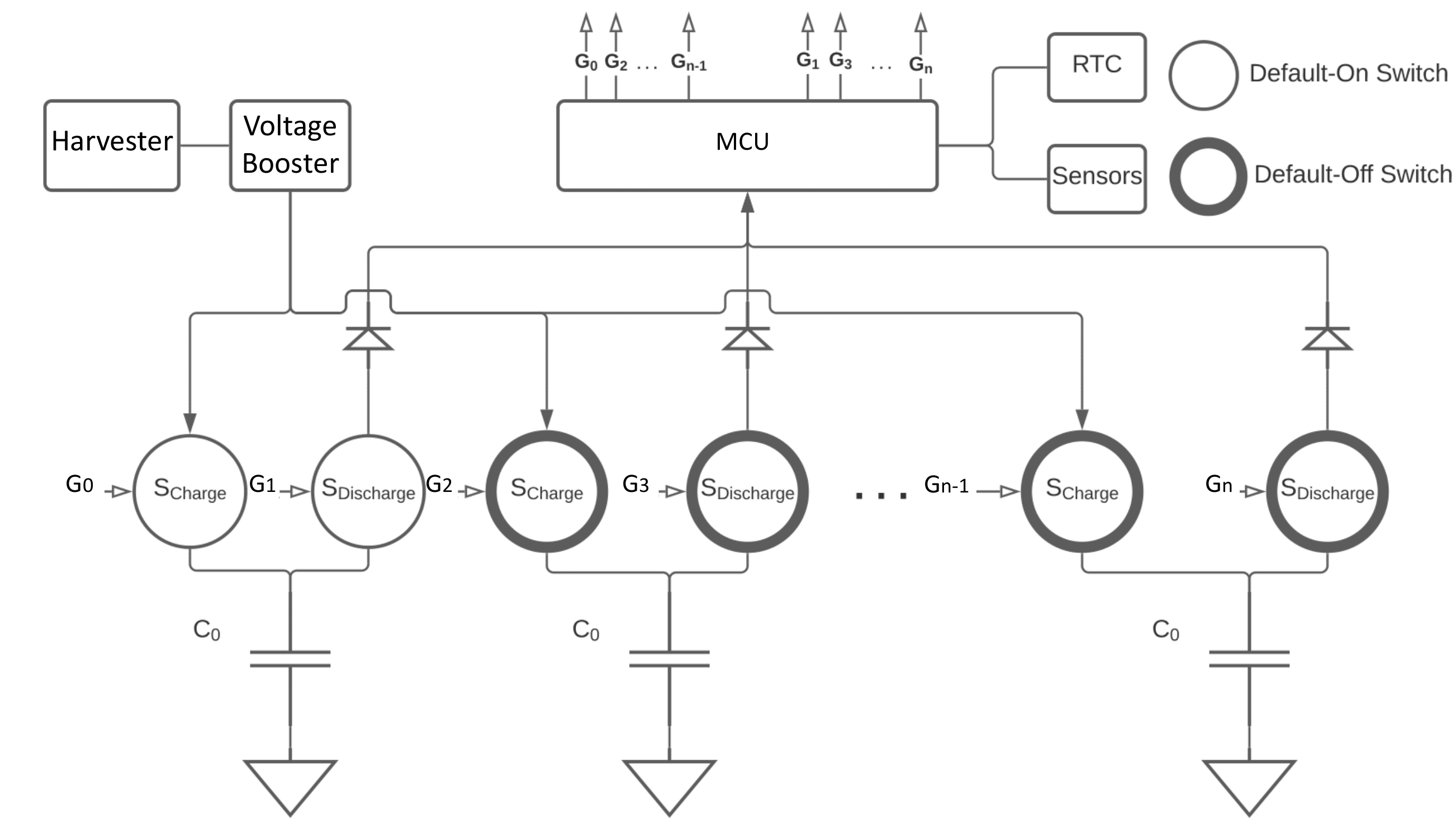}
    \caption{Circuit design of software-controlled cascading capacitor array}
    \label{fig:falinks_hardware}
\end{figure}

Figure~\ref{fig:falinks_hardware} shows the circuit design of the software-controlled cascading capacitor array. Each supercapacitor of the cascading array has two switches~\cite{smarton,luo2023efficient} connected to the harvester and the microcontroller that control capacitor charging (harvesting) and discharging (computing) at any time. When the system restarts, we use a default-OFF switch that disconnects the supercapacitor from the harvester/microcontroller. The first capacitor has a default-ON switch to ensure that at least one capacitor is active. Along with the capacitor array, the platform has a Cascaded Hierarchical Remanence Timekeeper (CHRT)~\cite{dereliable} for timekeeping and uses a low-power microphone at an 8 kHz sampling rate. Figure~\ref{fig:falinks_pcb} shows our prototype board with four capacitors -- one default-ON and three default-OFF switches. We use three identical nodes for this experiment. 

\parlabel{Data Processing Pipeline}
After data acquisition, each intermittent node first converts the time domain data into the frequency domain by performing a 256-point Fast Fourier Transform and then infers a deep neural network-based acoustic event detector. This detector consists of one convolution layer, two fully connected layers, and max pool and batch normalization. The mean inference time of this acoustic event detector in an MSP430FR5994 is 3.89 seconds, consuming 26.72 mJ energy on average. Note that we train this network using a regular GPU machine, then perform pruning to fit the network into the 256 KB non-volatile memory of the system. Finally, we convert the weights and program into suitable binary code to execute in the MSP430 microcontroller. 

\parlabel{Power Failure Protection}
Due to the limited harvestable energy and energy buffer, the entire data processing pipeline can only be executed with interruption. Besides, the stochastic nature of the energy sources causes random power failure. 
To ensure the forward progress of code execution and avoid corrupted results, we divide the pipeline into atomically executable units — which guarantees correct intermittent execution using SONIC API~\cite{gobieski2019intelligence}. SONIC stores important program states in the non-volatile memory after the execution of each unit. The system executes the next unit or re-executes the current unit (if it fails to finish before power failure) after each failure. Moreover, we utilize a double-buffered to ensure data integrity when the power supply is interrupted. We further reduce context switching and read-write overheads and minimize memory requirements by double-buffering~\cite{islam2020zygarde,zhang2022demo}.

\parlabel{Energy Source}
In this experiment, we only consider the two types of target energy source behaviors -- available energy is constant over time, and the available energy varies over time. For the fair evaluation of different algorithms, we use a Raspberry Pi logger with an Arduino Uno to collect solar and RF energy traces. We measure the voltage across the load resistor connected to the harvester and collect a total of 11.5 hours long trace. 

We use an Ethylene Tetrafluoroethylene (ETFE) based solar panel with an LTC3105 step-up regulator in indoor and outdoor scenarios. For outdoor scenarios, we deploy the energy collection setups on the pavement of a busy road where passing pedestrians and vehicles are blocking the sun and introducing unexpected unavailability of energy. The collected solar harvester collects up to 100mW of energy. 

We use the Powercast harvester-transmitter pair~\cite{powercast, powercasetransmitter} operating at 915 MHz to harvest RF energy (58-80mW). These energy traces contain the uncontrolled movement of the source and passing traffic (people walking between the harvester and the transmitter). Then we utilize these traces to execute different algorithms in the same energy condition.

\parlabel{Acoustic Event}
We use the Detection and Classification of Acoustic Scenes and Events (DCASE) 2018 challenge "\textit{Large-scale weakly labeled semi-supervised sound event detection in domestic environments}" dataset. It contains 3328 occurrences of ten acoustic events-- alarm bell ringing, dogs, cats, dishes, speech, frying, running water, blender, vacuum, and electric shaver.
An event's maximum and minimum duration is 10 and 0.25 seconds, and multiple events can coincide. To reduce the concurrent occurrence of multiple events, we only consider six types of events whose average duration is below five seconds. This subset has 3003 events with 307 alarm bells ringing, 450 dogs, 246 cats, 445 dishes, 1499 speeches, and 56 blender audios. 

In summary, we play six acoustic events, each lasting between 0.25 and 5 seconds. This dataset provides us with the event occurrence time, allowing us to evaluate different algorithms fairly.

\parlabel{Reward for Suboptimal Reinforcement Learning (SRL)}
The positive reward is straightforward and is given when a device can successfully capture an event. The negative reward is challenging because the device needs to know what happened during the power-off period. To determine the reward without additional sensing, we assume that it missed the data during the last cycle when a node fails to capture the beginning of an event (by observing a higher magnitude after a silence). Then, the system gets a random negative reward where the randomness ensures that not all nodes get the same negative rewards, which can lead to non-collaborative behavior. Note that the system knows the average duration of events as a priori. 


\begin{figure}[!htb]
    \centering
    \includegraphics[width=0.45\textwidth]{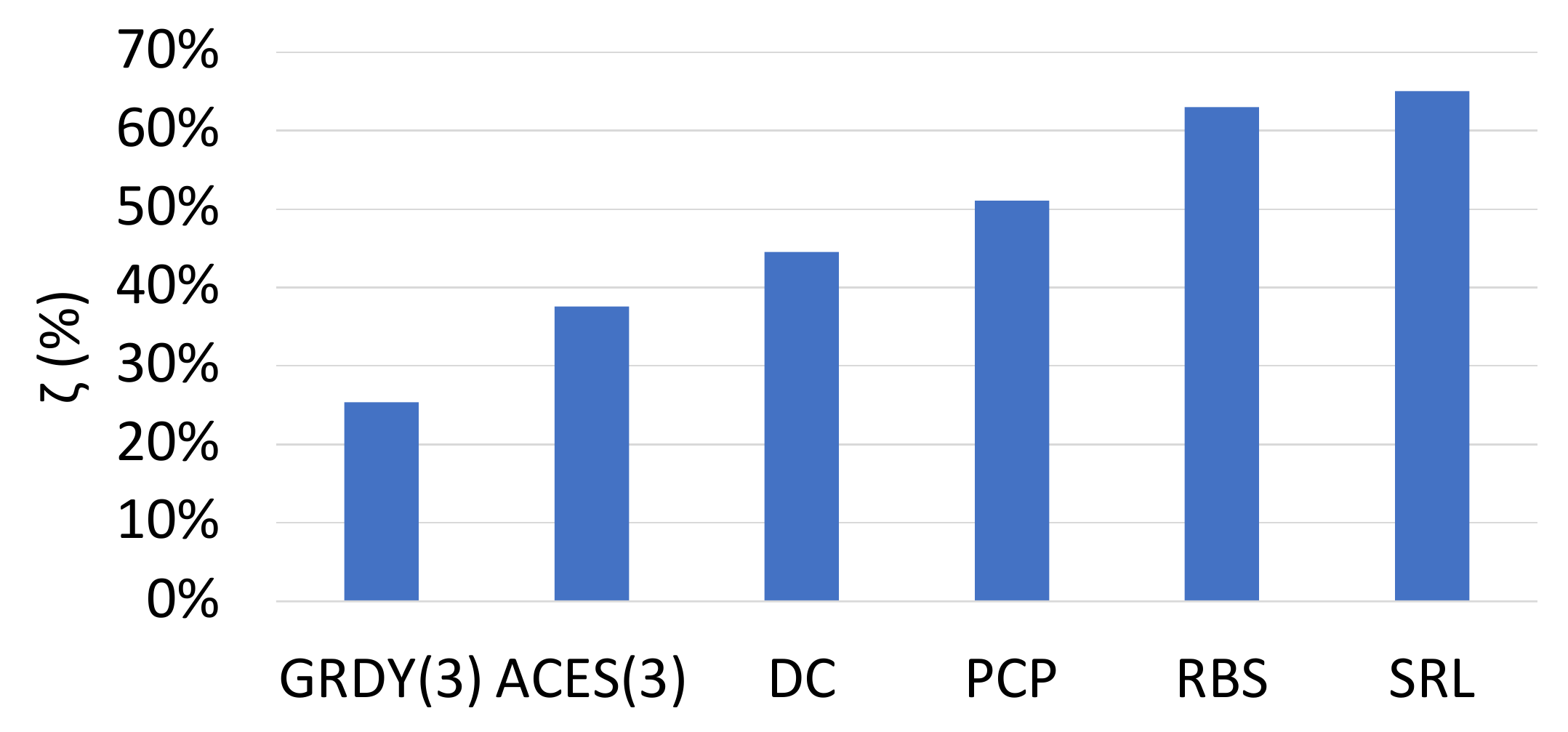}
    \caption{The capture and process success rate for various algorithms in real-world scenarios.}
    \label{fig:real_cpsr}
\end{figure}
\subsection{Performance Analysis}
Figure~\ref{fig:real_cpsr} provides the capture and process success rate comparison of our proposed algorithm with the baseline algorithms. As it is only possible to know every node's available and stored energy with communication overhead, we do not consider the Oracle baseline in this real-world experiment. We observe that the greedy swarm, GRDY(3), and the ACES swarm, ACES(3), reflect the $\zeta$ of the simulation during the real-world experiment. Compared to the simulated scenario, the capture and process success rate of fixed and general duty cycles (DC) drops by almost 20\% in the real-world scenario. The simulated result is an average over hundreds of energy scenarios, while the real-world experiment is close to 12 hours of solar and RF traces. Our investigation reveals that in some energy conditions during the simulation, the capture and process success rate is higher than 80\%. In comparison, in some other cases, it is lower than 30\% bringing our average close to 64\%. Moreover, the real-world energy trace combines the constant and variable available energy over time, which we have no control over. 

This real-world experiment also reflects our previous observation that PCP fails to utilize the swarm of nodes fully. Though SRL performed better than RSB in the simulation, in the real world, RBS and SRL can successfully capture and process a similar (less than 3\% difference) number of events. Though these two algorithms' average capture and process success rate are 65\%, they only missed 24.91\% of events on average. The lower $\zeta$ occurred due to failure to complete jobs within the deadline. RBS performs better than simulation because the energy is less random in the real world than in the simulation. 

\begin{figure}[!htb]
    \centering
    \includegraphics[width=0.45\textwidth]{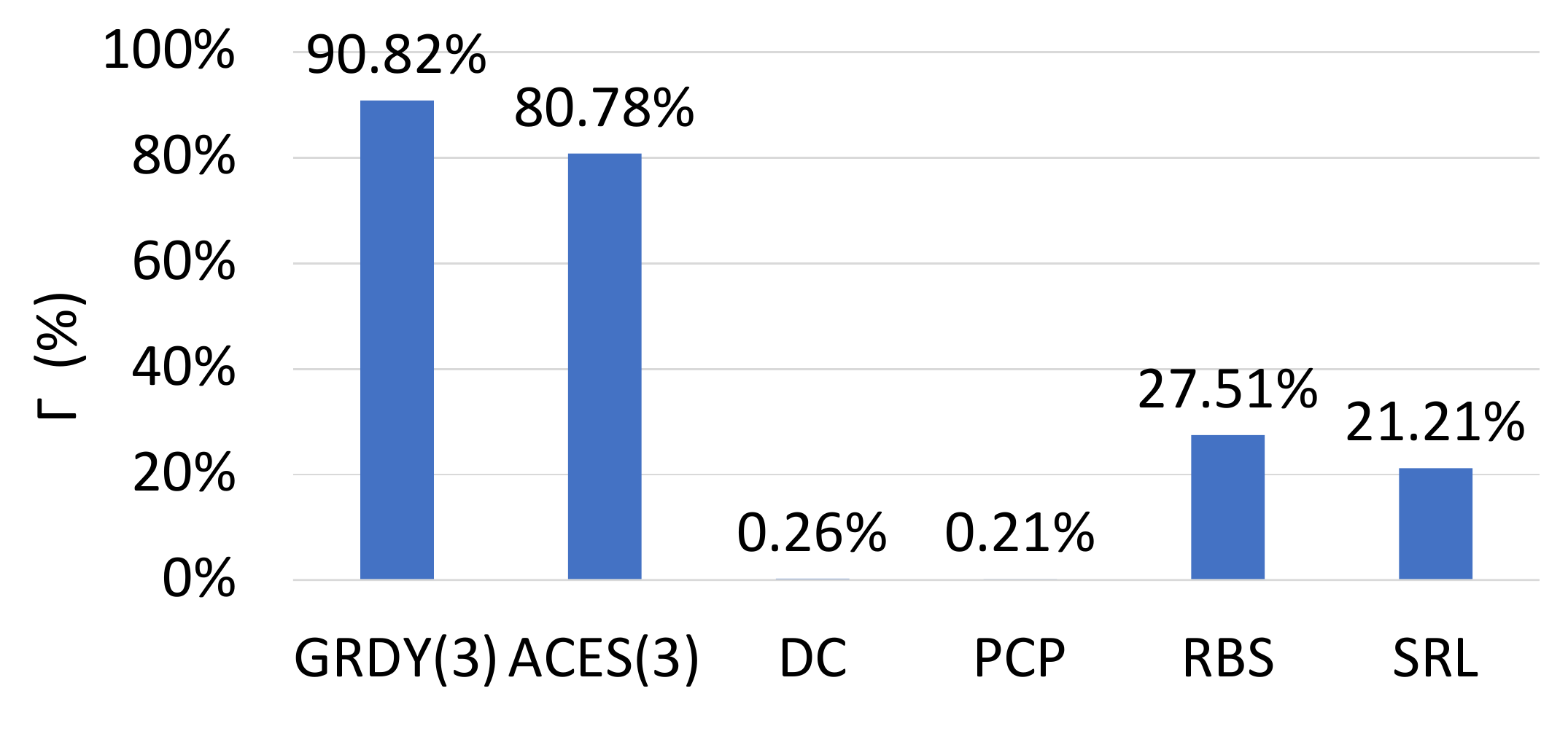}
    \caption{Redundant active time for various algorithms in real-world scenarios.}
    \label{fig:real_rat}
\end{figure}

Figure~\ref{fig:real_rat} shows the redundant active time in the real-world scenario. More than one node has the same active time more than 90\% of the time, which is more than what we have observed in our simulated experiments. Similarly, ACES also observes higher redundant active time than the simulation. The less aggressive clock drift in the real world than in the simulation may contribute to this. Though RBS and SRL have similar capture and process success rate, SRL has lower redundant active time due to the more tailored adaptation by this heuristic. Despite the higher active redundant time than simulation, SRL has higher capture and process success rate than other algorithms. 
\section{Related Works}
This section describes the existing literature on intermittent computing, scheduling battery-free systems, and distributed sensor systems.

\parlabel{Intermittent Computing}
Due to the stochastic nature and the lack of adequate power from the energy sources, batter-free systems experience frequent power failures. These failures result in the re-execution of instructions and inconsistency in non-volatile memory. For reliable forward execution of instructions despite power failure, previous works use software check-pointing~\cite{ransford2012mementos, maeng2018adaptive, hicks2017clank, lucia2015simpler, van2016intermittent, jayakumar2014quickrecall, mirhoseini2013automated, bhatti2016efficient}, hardware interruption~\cite{balsamo2015hibernus, balsamo2016hibernus++, mirhoseini2013idetic}, atomic task-based model~\cite{maeng2017alpaca, colin2016chain, colin2018termination}, and non-volatile processors(NVP)~\cite{ma2017incidental, ma2015architecture}. Some recent works have explicitly focused on inferring deep neural networks in batteryless systems by combining atomic task-based models with loop continuation~\cite{gobieskiintermittent, gobieski2019intelligence} or introducing imprecise computing~\cite{islam2020zygarde}. We utilize the SONIC~\cite{gobieskiintermittent, gobieski2019intelligence} task-based execution model to implement deep neural network inference. Recent research on intermittent computing focuses on formally verifying the systems~\cite{bohrer2022batfly} and designing developer-friendly platforms~\cite{bakar2022protean}. However, none of these works focus on validating whether the task can be finished or have considered multiple intermittent nodes. 

\parlabel{Scheduling in Intermittent Computing Systems}
Prior works have explored optimum voltage calculation~\cite{buettner2011dewdrop}, execution rate adaptation~\cite{sorber2007eon, dudani2002energy}, and stale data discarding to increase task completion likelihood~\cite{hester2017timely}. Some recent works have proposed scheduling algorithms for intermittent computing tasks to address the missing part of deadline-aware execution~\cite{islam2020zygarde, islam2020scheduling, yildirim2018ink}. None of these works consider the potential of scheduling multiple nodes to utilize their collaborative potential fully. Recently some works for intermittent sensor systems use reinforcement learning to increase the performance of intermittent computing nodes~\cite{fraternali2020aces, smarton}. None of these works consider the potential of using multiple nodes. 

\parlabel{Distributed Sensor Systems}
Though literature exists on distributed wireless sensors~\cite{zheng2013survey, zhu2019broadcast}, these algorithms cannot be directly applied for collaborative intermittently-powered systems as they are assumed to have communication among them. Communication is the most energy-expensive operation, and such frequent communication lowers the utilization of the system. 

Though recent works~\cite{torrisi2020zero, geissdoerfer2021bootstrapping} have shown the possibility of using zero-power backscatter communication, they can only communicate with one node at a time, thus not sufficient for emulating always-on scenarios. Recent work~\cite{geissdoerfer2021bootstrapping} uses external catalysts like light flickering, which is not generalizable. 
Moreover, active radio requires the transmitting and receiving battery-free node to turn on the radio of both devices at the same time. Recent works~\cite{geissdoerfer2022learning} have proposed wake-up schedules and neighbor discovery for effective communication in an intermittent computing system. However, they only support communication with one node and a known energy pattern. For a swarm of identical nodes to synchronize, multiple-node support is necessary, which involves the recovery of conflicting packets. 
Our closest related work, CIS~\cite{majid2020continuous}, enables continuous sensing with distributed battery-free devices. Unlike \Sys, CIS only considers the scenario where the energy available to every node is the same.

\section{Discussion}

\parlabel{Handling Different Events}
Though this paper assumes that all nodes capture all events equally (global), many real-world events are sensed differently at each node (local). For example, all swarm nodes in a building can hear a fire alarm, which is a global event. On the contrary, only some of the sensors around a building can sense when a person is breathing irregularly. Note that only global events are in the scope of this paper. 

When events are local, a subset of intermittent nodes can sense the target event. This subset selection can be formulated as a wireless sensor network formation problem, such as a topology-based network with cluster-based formation. Here the cluster formation is done based on the predetermined subset of the nodes for the event type. A single node can belong to multiple clusters, and after formulating the clusters, our proposed Falinks algorithms apply to each cluster. 

It is more beneficial to select the nodes belonging to the maximum number of clusters to select the minimum number of nodes while having at least one node from each cluster. In summary, along with the Prime-Co-Prime algorithm, which uses the node with the highest energy harvesting rate, using the node that belongs to most clusters more frequently is effective.

\noindent\textbf{Why not compare with Coalesced Intermittent Sensor (\textbf{CIS})?}
CIS~\cite{majid2020continuous} only works for scenarios where the target energy is variable but equal for every node. Here, each node keeps track of the energy at the other nodes and then calculates the number of active nodes to randomly chooses whether to sleep or wake up. We do not compare \Sys with CIS, as it fails to precisely count active nodes when the available energy at each node varies, and \Sys exclusively focuses on scenarios where available energy is different at each node. 

\parlabel{Dependency on Super-capacitor Discharging}
Most battery-free nodes use supercapacitors as energy buffers. Unlike batteries, supercapacitors have high charging and discharging rate. Therefore, even if a node does not utilize its energy greedily, it may still lose unused energy due to self-discharging. However, as other nodes are already active while a node is preserving energy, the swarm's performance remains unaffected. However, due to the sub-optimal heuristics, this energy loss might lower the performance when the available energy source varies across time and at every node. Therefore, a lower discharging rate is beneficial for our proposed algorithm.

\parlabel{Proof of Optimality of the Oracle Algorithm}
The following proof shows that -- "\textit{when the available energy at all nodes is known, waking up only the node with the maximum total harvestable and stored energy is optimal}."

\itparlabel{Proof} 
To prove this theorem using contradiction, we assume that an intermittent node $I_i$ ($i = 1, 2, ..., N$) has total energy $E_{i} = E_{H_i} + E_{C_i}$, where $E_{H_i}$ and $E_{C_i}$ are the available energy to harvest and energy stored for the $i^{th}$ intermittent node. Let $I_m$ be the node with maximum total energy, $E_m = max(E_i)$, for $\forall i \in N$. Let us assume that waking $I_m$ is not optimal. Thus, another node $I_p$ exists with $E_{min} < E_p < E_m$, and waking $I_p$ is optimal. Here, $E_{min}$ is the minimum operational energy. 

$E_m > E_p$ can happen if $E_{H_m} > E_{H_p}$ or $E_{C_m} > E_{C_p}$. If $E_{H_m} > E_{H_p}$ and we decide to wake only $I_p$ up while keeping all the other nodes asleep, ${E_p}' = E_p - E_{min} + E_{H_p}$ and ${E_m}' = E_m + E_{H_m}$. As  $E_{H_m} > E_{H_p}$ and ${E_m}' > {E_p}'$, $E_m$ will keep increasing and reach the maximum energy storage capacity. Then, $I_m$ will fail to harvest available energy, wasting potential energy. Similarly, when $E_{C_m} > E_{C_p}$, the energy storage of $I_m$ will fill up more quickly, wasting available energy. As wastage of potential energy is never optimal, waking up $I_p$ is not optimal, which contradicts our assumption. Therefore, we prove that waking $I_m$ up is optimal. \proved

\parlabel{Position of the Intermittent Nodes}
The placement of intermittent nodes is essential to avoid missing an event. A sensor swarm can miss sensing an event if (1) all of them are inactive or (2) the event is out of all nodes' sensing range. Thus the placement of intermittent nodes depends on three factors -- energy source, event source, and physical constraints. This paper focuses on finding the appropriate duty cycle of pre-positioned intermittent nodes instead of looking at placing the nodes. However, the PrimeCoPrime algorithm can be used to determine the placement with respect to the energy sources. The placement concerning the event source can be formed as the art gallery problem~\cite{abrahamsen2018art}, a well-studied visibility problem in computational geometry. Combining all three factors for determining a node's placement is an unsolved problem that we will address in future works.
\section{Conclusion}

In this paper, we study the unique problem of intermittently powered systems, where the system fails to observe or sense the target event due to insufficient energy to turn on. We take a unique approach to the problem by considering a swarm of intermittent nodes as an entity that amalgamates to address this non-observability problem. However, the high communication cost hinders this collaborative behavior by imposing no communication rules for efficiency. Therefore, we propose \Sys, where a group of intermittent nodes independently wakes up and sleeps to increase their collaborative capture and process success rate while minimizing the redundant active time by keeping at least one node active most of the time. On average, it captures and processes 58.50\% more events successfully than a single intermittent node and has a 54.40\% more capture and process success rate than a non-amalgamated swarm of intermittent nodes.

\begin{acks}
We want to thank Luca Motolla for giving suggestions on the manuscript.
This paper was supported, in part, by NSF grants CNS 2047461. 
\end{acks}

\bibliographystyle{ACM-Reference-Format}
\bibliography{falinks}

\end{document}